\documentclass[12pt]{article}

\begin{document}

\markboth{Aldo L. Cotrone}
{YM and QCD spectra from five dimensional strings}

%
%
\begin{center}
{\LARGE{On the YM and QCD spectra\\
from five dimensional strings
}}
\end{center}

\vskip 10pt
\begin{center}
{\large Aldo L. Cotrone}
\end{center}

\vskip 10pt

%

\begin{center}
\textit{Departament ECM, Facultat de F\'isica, Universitat de Barcelona and \\Institut 
de Fisica d'Altes Energies,\\
Diagonal 647, E-08028 Barcelona, Spain.\\
and\\
Institute of Theoretical Physics, KU Leuven\\
Celestijnenlaan 200D, B-3001 Leuven, Belgium.\\
aldo.cotrone@fys.kuleuven.be}
\end{center}



\begin{abstract}
We consider a non-critical five dimensional string setup which could provide a dual description of QCD in the limit of large number of colors and flavors. The model corresponds to $N_c$ color D3-branes and $N_f$ D4/anti D4-brane pairs supporting flavor degrees of freedom.
The matching of the string model spectrum with the dual field theory one is considered.
We discuss the consequences of the possible matching of the gravity modes with the light glueballs and the interpretation of the brane spectrum in Yang-Mills and QCD.

\end{abstract}


\section{Introduction}\label{secintro}

Planar four dimensional YM theory and QCD may have a description as string theories in at least five dimensions \cite{thooft,veneziano,polyakov}.
We will follow the prejudice that such a description exists and that five dimensions are sufficient. 
Such a dual to YM or QCD must have a background with curvature of the geometry of order one in $\alpha'$ units.
Thus, a gravitational description of the system is not a good approximation and we should really solve the associated sigma-model.
Of course, at present this is an impossible task.
The best we can do is to try and learn some qualitative features of these theories, using the minimal amount of assumptions we can formulate about the actual string theory dual of YM or QCD.\footnote{Nevertheless, solving the 5d gravity equations, although being not a fully consistent procedure, can in principle give important qualitative indications about the sigma model \cite{wall}.}

Besides defining better the problem, this exercise could provide a more stringy-inspired framework for the phenomenological models that go under the name of AdS/QCD \cite{katz,pomarol}.
For example, one often encounters in these models some five dimensional lagrangians which include a scalar field dual to the quark bilinear, that have a natural derivation from the expansion of the DBI action for systems of branes/anti-branes supporting the flavor degrees of freedom, the quark bilinear being the open string tachyon.\footnote{As such, it could be worthwhile to include in these models the first subleading corrections coming from the DBI, which possibly would ameliorate the numerical results.}

In these notes, we will formulate the setup that we believe is the most natural candidate for describing YM and QCD in five dimensions, essentially borrowing all the data from our knowledge of the ten dimensional cases \cite{tutti1,tutti2,tutti3,tutti4}.
Clearly, since we have no control on the theory, some of the material will be speculative; nevertheless, we believe it is worthwhile to begin such a discussion.

In the next section we will describe the basic features of the 5d models, in order to settle the background for a qualitative comparison of (some part of the) spectra of string theory with the ones of YM and QCD.
The scalar and vectorial mesonic sector of the latter has recently been described in  \cite{poli,ckp}, so we will just briefly report the proposal at the end of section \ref{secsetup}.
Then, in the following sections, we will discuss the possible qualitative matching of the gravity and non-perturbative spectrum of string theory with the YM one.

Interestingly, despite our ignorance about strings in 5d, it seems that, assuming that the theory is not very different from the known 10d ones, we can have a (almost) consistent picture of the spectrum of YM theory from the string.
Moreover, we can deduce some interesting possible implications for both theories.
For example, the glueball spectrum indicates that among the string fields there should not be any tachyon dual to a light glueball, nor any degenerate doubling of the RR spectrum typical of a Type 0 theory in 10d, as also argued recently in  \cite{Thorn:2008ay}.
Moreover, the consistency of the D-brane spectrum with the YM one indicates that the YM coupling is likely to blow up in the far IR, rather than relaxing to a constant value of order one.  

Unfortunately, we could not give a complete discussion of the spectrum in these notes.
Among the interesting features we do not discuss there are important issues such as the topological susceptibility and the gluon condensate.
We hope to be able to discuss these features in the future.

As a final remark, we report that in  \cite{gk,gkn} some of the issues discussed here are approached in the context of an effective gravity model in five dimensions, in the spirit of AdS/QCD. Our perspective has the complementary advantage of pointing out that some results do not rely on a specific gravity model (which is not fully under control as a string approximation in five dimensions).

\section{The setup}\label{secsetup}

The building blocks of YM and QCD are expected to be as follows.
The strongly curved five dimensional background dual to $SU(N_c)$ YM is generated by a stack of $N_c$ color D3-branes \cite{tutti1,tutti2,tutti3,tutti4,wall}.
It includes at least a running dilaton and a space-time filling five-form RR field $F_5$, proportional to the number of colors $N_c$.
There could also be a non-trivial axion turned on, giving a non-zero $\theta_{YM}$ angle.
The metric in string frame reads \footnote{For obvious reasons, we will not keep track of $2\pi$ and $\alpha'$ factors.}
\begin{equation}\label{metric}
ds^2 = e^{2f(r)}dx_\mu dx^\mu + dr^2\ ,
\end{equation}
where the warp factor for the Minkowski part of the metric, $e^{2f(r)}$, approaches a non-vanishing constant at the bottom of the space (this is necessary for confinement \cite{wym,kinar}), that we will put at $r=0$.
This corresponds to the IR of the dual field theory.
At large radius, that is in the UV of the field theory, the metric should approach $AdS_5$, $e^{2f(r)}\sim e^{2 k r}$ (with radius $R^2=1/k^2$ of order one in $\alpha'$ units), since we require asymptotic freedom; nevertheless, we are not going to use this assumption.   

Since pure YM theory has no space-time fermions, while presumably the string theory consistency requires world-sheet supersymmetry \cite{wall}, the dual string theory could be similar to the ten dimensional Type 0.
The latter is known to contain a closed string tachyon $T$ and his spectrum of RR fields is doubled: there are two sets of Dp-branes, named ``electric'' and ``magnetic'', for every p \cite{kt1}.
It is not clear if the doubling of the RR spectrum persists in 5d \cite{armoni,ferretti2,Thorn:2008ay}.
If the doubling survives in 5d, we have to see the background above as generated by one type of D3, say the ``electric'' ones.
Finally, the tachyon $T$ could be possibly non-trivial \cite{armoni,ghor1}.

On this background, the glueballs are described by the perturbative string states, while the D-brane spectrum should be matched with other field theory degrees of freedom, as we will try and do in the following sections.

In order to describe QCD, we must include fundamental fermions.
Following the proposal in   \cite{km} for the six dimensional dual of SQCD, it was proposed in  \cite{poli} that in QCD the $N_f$ fundamental and anti-fundamental quarks are described by $N_f$ pairs of space-time filling D4/anti D4-branes.
If the theory is some kind of Type 0, in the spectrum of the strings connecting ``electric'' D3 and ``magnetic'' D4-branes there should be only fundamental fermions \cite{kt1,kt3,costa}.
Instead, in the spectrum of strings with both end-points on the D4-branes (the mesons) there should be only bosons (that is, there should not be ``fermionic mesons'' as in the usual 10d models).

The two groups of branes/anti-branes account naturally for the flavor symmetry $U(N_f)\times U(N_f)$ prior to chiral symmetry breaking.
The latter is triggered by the quark condensate, that is a vev for the scalar field in the bifundamental of the two groups (the pion).
In the string description the chiral condensate is nothing else that a non-trivial profile for the open string tachyon $\tau$ in the brane/anti-brane system \cite{sugimoto,KarKatz}.\footnote{The idea that the tachyon should be the pion was already present in the early seventies \cite{schwarz}.}   
Its presence in fact breaks the flavor group to a diagonal subgroup,  $U(N_f)\times U(N_f) \rightarrow SU(N_{f})_D \times U(1)_B \times U(1)_A$, where $U(1)_B$ is of course the baryonic $U(1)$ and the axial $U(1)_A$ is anomalous \cite{ckp}.

The scalar and vector meson spectrum is naturally accounted for by the open string tachyon $\tau$ and the gauge fields on the D4/anti D4-brane world-volume \cite{poli}.
The complex tachyon gives a full tower of scalar and pseudo-scalar mesons, comprising the Goldstone pions, while the world-volume gauge fields account for a full tower of vector and axial-vector mesons.
This was shown in  \cite{ckp}, where it was pointed out that this picture is consistent provided that the open string tachyon $\tau$ is diverging at the bottom of the space $r=0$, and that this condition can give ``linear confinement'' for highly excited mesons.
Finally, the quark mass is introduced by turning on the non-normalizable mode of $\tau$ at large radius. 
This part of the QCD spectrum is reported in Table \ref{qcdmesons}.
\begin{table}[ht]
\caption{Matching of the scalar and vector QCD mesons with D4/anti D4 world-volume modes.}
\begin{center}
{\begin{tabular}{@{}ccccc@{}}\hline
$J^{PC}$: & $0^{++}$ & $0^{-+}$ & $1^{--}$ & $1^{++}$ \\
\hline
DBI mode: & $\tau+\tau^\dag$ & $i(\tau-\tau^\dag)$ & $\frac12 (A_L+A_R)$ & $\frac12 (A_L-A_R)$ \\
\hline
\end{tabular}}
\label{qcdmesons}
\end{center}
\end{table}
In the table, $A_L, A_R$ are the gauge fields living on the world-volume of the $N_f$ flavor branes and anti-branes respectively (since the branes are space-time filling, there are no scalar fields on the world-volume).
The pseudo-scalars $0^{-+}$ are of course the pions.
We are not going to discuss the flavor sector anymore in these notes, even if we lack some basic ingredients (for example, the dual to the mixed quark-gluon condensate).

\section{YM spectrum matching}

Having reported the recent results on the meson spectrum in the previous section, here we analyze the pure glue $SU(N_c)$ YM theory.
Besides the assumptions on the form of the YM dual formulated above, here we will add the following:
first, even if gravity is not a good approximation to the theory, and so the {\emph{usual}} gravity modes are not much lighter than the other string states, we will reasonably assume that they are still among the lighter modes in the true string spectrum; second, we will assume that if the RR spectrum is doubled, one of the two RR fields of each type is much heavier than the other.

Concerning the latter assumption, we will encounter no indication for a doubling of the RR (and brane) spectrum.
Moreover, even if the spectrum is doubled, the properties of the ``electric'' and ``magnetic'' objects would be different, since the background is generated by ``electric'' branes only.
Thus, we would expect anyway a non-degenerate mass spectrum.
On the other hand, we notice that at least some of the 10d gravity duals of YM-like theories, such as the ones in  \cite{wym,min2}, have a ``doubled'' brane spectrum in 5d.
In fact, considering also the branes wrapped on the internal cycles, they have two kinds of branes of the same dimensionality in 5d.
For example, in the Type IIA model of  \cite{wym}, besides the ordinary D0-branes, one has D4-branes wrapped on the internal $S^4$, that are ``D0-like'' in five dimensions; analogously, besides the D2-branes, one has wrapped D6, and so on.
Note that these branes of the same dimension in 5d have different physical properties, which would be the case also in the present setup because of their ``electric'' or ``magnetic'' nature. 
It would be interesting to understand whether this apparently frequent ``doubling'' of the brane spectrum in the duals of YM-like theories has some real implication for the actual YM theory.

Anyway, in the following we will not consider the possibility of this ``doubling''. 
We will begin by the matching of the gravity spectrum and then come to the matching of the non-perturbative string spectrum of branes.
We will adapt ten dimensional techniques to the five dimensional case.
This naive procedure provides a somewhat consistent picture.

\subsection{Glueballs vs. gravity spectrum}

The perturbative string spectrum should be matched with the glueball spectrum of YM.
As usual, almost every glueball is dual to a genuinely string state.
If nevertheless we insist on the assumption that the usual gravity modes are among the lighter ones in the spectrum, we should be able to match their spectrum with the lighter glueballs.
The latter are known from lattice simulations.
Since we do not have any computable string model at hand, we do not compare masses,\footnote{A precise matching of the masses between lattice and gravity data is not expected even in 10d, being the two computations done in different regimes of parameters \cite{brower}.} but just search for the qualitative matching of the gravity and lattice modes. 

\subsubsection{Lattice data}
We use the data on the $SU(8)$ theory (the nearest to the large $N_c$ limit), that can be found in  \cite{mt,meyer}.
We report the relevant data for the continuum spectrum in Table \ref{glueballs}.
\begin{table}[ht]
\begin{center}
\caption{Continuum $SU(8)$ glueball spectrum.}
{\begin{tabular}{@{}cccccccc@{}}\hline
$J^{PC}$: & $0^{++}$  & $2^{++}$ & $0^{++*}$ & $0^{-+}$ & $0^{++**}$ & $2^{-+}$ & $1^{+-}$ \\
\hline
$m/\sqrt{\sigma}$: & 3.32(15) & 4.65(19) & 4.71(29) & 4.72(32) & $\sim $ & 5.67(40) & 5.70(29) \\
\hline
\end{tabular}}
\label{glueballs}
\end{center}
\end{table}
In the table, the masses are normalized with respect to the string tension $\sigma$.
The reported masses are the average ones.
We include only the lightest modes up to the first glueball that is surely dual to a string mode, namely the $2^{-+}$: the only spin two field in the gravity spectrum is the graviton, dual to the $2^{++}$ glueball.
As an exception, we also report the $1^{+-}$ glueball, since the errors are such that it could be lighter than the $2^{-+}$.
In any case, since the sting theory is at strong curvature, the string modes will not be much heavier than the gravity ones, consistently with the actual glueball spectrum.
Of course, it can well be that some stringy modes are lighter than some of the usual gravity modes.
In fact, this is what most probably happens.
We are nevertheless interested in matching only the gravity modes with the appropriate glueball states.

The first glueball we do not report is the $2^{++*}$ excitation, with mass around $6.5$.
We instead include the $0^{++**}$, since it appears in the table 7.15 of  \cite{meyer}, even if it is not reported in the continuum spectrum.
Its mass is slightly above the $0^{-+}$ one.

As a general remark, the $SU(3)$ spectrum is very similar to the $SU(8)$ one, but for the masses of the $0^{++*}$ and $0^{++**}$, which are much heavier.
In particular, the first one is heavier than the $1^{+-}$.
As such, these $0^{++}$ states could be peculiar to the large $N_c$ limit, that is, they could be the first excitations of k-strings \cite{meyer}; their masses fit well this explanation. 
We will discuss this point further in the following.

\subsubsection{Gravity modes}\label{secgravity}

In 5d the gravity modes have been argued in  \cite{kt1,ferretti2} to include: the graviton $g$ (with five physical degrees of freedom), the dilaton $\Phi$, the NSNS two-form potential $B$ which is dual to a one-form potential $A_1^{NS}$ (with three physical degrees of freedom), and possibly a real closed string tachyon $T$ from the NSNS sector; the axion $\chi$, a one-form potential $A_1^{R}$ dual to a two-form potential $C_2$, and some fields with no propagating degrees of freedom (as it is evident considering the duals) as the four-form potential $C_4$ and the five-form potential $C_5$, in the RR sector.
As said, we make the assumption that if the RR sector is doubled, one of the two fields of any kind is much heavier that the other, so we will not consider it.

The $P, C$ quantum numbers of the dual glueballs of these gravity fields can be inferred from the way the latter are coupled to the YM field in the DBI and WZ action for a probe brane \cite{ooguri,brower}.
In our case the probe brane is of course a D3, so the terms in the WZ action that contain the YM field $F$ are
\begin{equation}
S_{WZ} \sim \int d^4x\, (\chi \ F \wedge F + \chi \ F \wedge B + C_2 \wedge F )\ .
\end{equation}
Under a parity transformation the YM gauge field transforms as: $A_i\rightarrow -A_i$ for $i=1,2,3$ and $A_t\rightarrow A_t$.
Thus, every $F_{\mu\nu}$ field with both indexes in $x_{1,2,3}$ is parity even, while $F_{it}$ is parity odd.
Charge conjugation acts as: 
$A_\mu \rightarrow -A_\mu^T$ (in matrix notation).
Since for more than one $F_{\mu\nu}$ this implies that the order of the fields is reversed, it follows that symmetric products of an odd number of fields will have $C=-1$ (antisymmetric products of an odd number of fields will have $C=1$, but these do not enter the game).
On the other hand, products of an even number of fields will have $C=1$.

From this, the parity assignment of the fields follow.
Let us first consider the field $B$.
From the coupling $\chi F \wedge B$, since $F_{it}$ is parity odd and $\chi$ is odd too (see below), it follows that $B$ is even (here we use the gauge where the $t$ components of the gravity fields are zero).
Furthermore, from the expansion of the square root in the standard DBI action, one reads that the coupling of $B$ with $F$ is of the form $B_{\mu\nu}\cdot Sym[F_{\mu\nu}W]$, where $Sym$ stands for ``symmetric product'' and $W$ contains an even number of $F$ fields (it is present only in the non-abelian case).
Thus, one immediately reads that $B$ is parity even (note that here it has the same indexes as $F$) and that it is charge conjugation odd, since there is an odd number of $F$s.

Analogously, from the WZ non-abelian action it follows that the coupling of the two-form $C_2$ is $\epsilon^{ijkt}C_{ij}\cdot Sym[F_{kt}W]$, so that $C_2$ is parity odd and charge conjugation odd.
The axion is instead coupled like $\epsilon^{ijkt}\chi\cdot Sym[F_{ij}F_{kt}W]$ so it is P-odd and C-even.
The graviton, that couples to the energy-momentum tensor\footnote{The energy momentum tensor $T_{\mu\nu}$ is a symmetric and (classically) traceless operator. Under the spatial rotation group $SO(3)$, which is the little group of massive particles, it decomposes as a spin 2 tensor $T_{ij}-trace$, a vector $T_{ti}$ and a scalar $T_{tt}$ (due to the zero trace condition on $T_{\mu\nu}$, the trace of $T_{ij}$ is equal to $T_{tt}$). 
The vector operator $T_{ti}$ does not create any glueball due to the conservation law for $T_{\mu\nu}$ \cite{Jaffe:1985qp}.} as $g_{ij}\cdot T_{ij}= g_{ij}\cdot Tr(F_{ik}F_{j}^k)$, is P-even and C-even, as the dilaton, that couples as $\Phi\cdot TrF^2$.
The possible closed string tachyon $T$ would presumably mix with the dilaton and again would be P and C-even.\footnote{In the known ten dimensional models, the tachyon enters the DBI action schematically as $\int k(T) e^{-\Phi}\sqrt{det(g_{\mu\nu}+F_{\mu\nu})}$, where $k(T)\sim 1 \pm \frac14 T +... $ and the two signs correspond to the ``electric'' and ``magnetic'' cases \cite{kt1,Klebanov:1998yy,Garousi:1999fu}. 
In the ``dyonic'' case, where there are two gauge groups, the tachyon couples to the operator $TrF_1^2-TrF_2^2$ \cite{Klebanov:1999um}, so the quantum numbers are $++$.
Unfortunately we do not know the coupling in the non-critical setting.
If the tachyon is present in the spectrum and it couples in a way similar to the ``electric'' ten dimensional case, it would probably mix with the dilaton.
Then, one combination would couple to $TrF^2$ (with an abuse of language we are going to refer to it as ``dilaton'').
Concerning the other combination (which we are going to call ``tachyon''), either it does not couple to any local operator and, as advocated in  \cite{Grena:2000xw,Csaki:2006ji}, it is dual to renormalons; or, if it is dual to a local operator, the most probable thing to happen is that the latter gives a P and C-even assignment as in ten dimensions \cite{Klebanov:1999um}.
In any case, we are going to argue that there is no light glueball candidate to be dual to this scalar field.}
These results are summarized in Table \ref{gravity modes}.

\begin{table}[ht]
\caption{Gravity spectrum in 5d.}
\begin{center}
{\begin{tabular}{@{}ccccccc@{}}\hline
Field: & $\Phi$ & $g$ & $T$ & $\chi$ & $B$ & $C_2$ \\
\hline
$J^{PC}$: & $0^{++}$ & $2^{++}$ & $0^{++}$ & $0^{-+}$ & $1^{+-}$ & $1^{--}$ \\
\hline
Components: & 1 & 5 & 1 & 1 & 3 & 3 \\
\hline
\end{tabular}}
\label{gravity modes}
\end{center}
\end{table}

\subsubsection{(Partial) Matching}

By comparison of Table \ref{glueballs} and Table \ref{gravity modes}, some identifications of glueballs with gravity modes are straightforward.
As usual, the graviton is obviously the $2^{++}$, five-component object dual to a massive spin two glueball with the same quantum numbers.
Analogously, the identification of the axion with the $0^{-+}$ is straightforward.

Moreover, using the duals of the two-forms above, the three component $A_1^{NS}$ and $A_1^{R}$ one-forms should be dual to the massive vectorial $1^{+-}$ and $1^{--}$ states.
Clearly, the $A_1^{R}$ one-form is peculiar.
The point is that the $1^{--}$ glueball is much heavier than the ones reported in Table \ref{glueballs} (its mass is around 7.5), so there are string modes lighter than this gravity mode.
In fact, this is not really a serious problem, as we know that the string modes are not separated from the gravity ones, and this mixing in the spectra is expected. 
For example, the $2^{-+}$ glueball is clearly a string mode.

Something interesting happens in the scalar sector.
The dilaton\footnote{What we call dilaton in this section is actually the unique scalar fluctuation diff. invariant, which is a combination of the dilaton and the metric, see for example  \cite{Kiritsis:2006ua}.\label{footnotedil}} and tachyon have the quantum numbers of the $0^{++}$ glueballs.
Nevertheless, in the light glueball spectrum there are three $0^{++}$ states and we do not have any gravity candidate for the dual to $0^{++**}$.
One possibility is that there is another scalar in the 5d gravity spectrum that we are overlooking.
Or, it could be that the third $0^{++}$ glueball is accounted for by a higher mode of the dilaton or the tachyon.
We believe this to be unlikely, for the following reason. 

The mass ratios of the $0^{++*}$ and $0^{++**}$ to the $0^{++}$ for $SU(8)$ on the lattice are about $1.42$ and $1.42-1.60$ (the uncertainty here comes from the lacking of the continuum mass value for the $0^{++**}$). 
In the $SU(3)$ case they are instead $1.87$ and $2.29$. 
As a comparison, the $0^{-+}$ to  $0^{++}$ ratios in the two cases ($SU(8)$ and $SU(3)$) are $1.42$ and $1.53$; the $2^{++}$ to the $0^{++}$ ones are $1.40$ and $1.46$;  the $1^{+-}$ to the $0^{++}$ ones are $1.72$ and $1.81$;
the differences in the masses between these $SU(8)$ and $SU(3)$ glueballs are about $10\%$ or less.
As such, it is not very likely that the $0^{++**}$ in $SU(8)$ could be the same state as the $0^{++*}$ in $SU(3)$, as the change in the mass ratio is from $1.42-1.60$ to $1.87$, an increase of at least the $17\%$ or more.
Analogously, it is also unlikely that the $0^{++*}$ in $SU(8)$ could be the same state as the $0^{++*}$ in $SU(3)$: since the mass of the $0^{++*}$ in $SU(3)$ is 1.8 times that of the $0^{++}$, we expect the corresponding state in $SU(8)$ to have a mass of order 6.2, well above the $0^{++*}$, $0^{++**}$ states measured on the lattice for $SU(8)$. 

Thus, as observed in  \cite{meyer}, it seems that {\emph{both}} the $0^{++*}$ and $0^{++**}$ are more likely to be k-string excitations rather than ordinary gravity modes.
As such, we are led to conclude that there is no other scalar beyond the dilaton in the gravity sector, dual to light glueballs.
That is, there should be no ``light'' tachyon.
To be more precise, the statement is that even if a tachyon exists in the spectrum of the string theory, out of the two scalars that come out from its possible mixing with the dilaton (and the metric, see footnote \ref{footnotedil}), one will be dual to a much heavier glueball than the other. 
We will continue to call ``dilaton'' the ``lighter'' of these two modes.\footnote{As already said, there is also the possibility, advocated in  \cite{Grena:2000xw,Csaki:2006ji}, that the tachyon is dual to renormalons.}
For similar reasons, we conclude that the RR fields which are the possible Type 0 doubles of the ones considered here, should be dual to much heavier glueballs, since there is no sign of any doubling in the lattice spectrum.

We put the proposed identification of the glueball spectrum in Table \ref{identification},\footnote{Note that in QCD, besides the $0^{++}$ corresponding to $Tr F^2$, there are two other $0^{++}$ states with corresponding operators of dimension four (the energy density) and six ($f_{abc}F^{a}_{\mu\alpha}F^{b\alpha\beta}F^{c\mu}_{\beta}$), see for example  \cite{brower}. The claim is that these glueballs are heavier than the ones reported in Table \ref{glueballs}.} together with the dimension $\Delta$ of some of the dual operators, as in  \cite{ooguri,brower}.
\begin{table}[ht]
\caption{Identification of the light glueball modes with string modes.}
\begin{center}
{\begin{tabular}{@{}cccc@{}}\hline
$J^{PC}\qquad $ & Field & Operator & $\Delta$ \\
\hline
$0^{++}\qquad $ & dilaton $\ \Phi$ & $\qquad Tr F^2\qquad $  & 4 \\
\hline
$2^{++}\qquad $ & metric $\ g$ & $\qquad T_{ij}-trace\qquad $ & 4\\
\hline
$0^{++*}\qquad $ & k-string mode &  &  \\
\hline
$0^{-+}\qquad $ & RR axion $\ \chi$ & $\qquad Tr F\wedge F\qquad $ & 4\\
\hline
$0^{++**}\qquad $ & k-string mode &  & \\
\hline
$2^{-+}\qquad $ & string mode & &  \\
\hline
$1^{+-}\qquad $ & NSNS one-form $\ A_1^{NS}$ & $\qquad d_{abc}F^{a}_{\mu\alpha}F^{b\alpha\beta}F^{c}_{\beta\nu}\qquad $ & 6\\
\hline
$1^{--}\qquad $ & RR one-form $\ A_1^{R}$ & $\qquad d_{abc}F^{a}_{\mu\alpha}F^{b\alpha\beta}F^{c}_{\beta\nu}\qquad $ & 6 \\
\hline
\end{tabular}}
\label{identification}
\end{center}
\end{table}
In the table, $T_{\mu\nu}$ is the dimension four stress-energy tensor of YM and $d_{abc}$ is a symmetric coupling.

\subsection{Non-perturbative string spectrum}

Apart from the glueballs, the YM spectrum contains other physical degrees of freedom.
Also, in string theory there are non perturbative states.
In this section we try and match the two spectra as much as possible.
We first consider the string theory and then argue to what YM degrees of freedom the various states could correspond to.

\subsubsection{Branes}

In string theory there are a collection of D-branes and a solitonic NS-brane.
As usual, p-branes couple to (p+1)-form potentials.
Thus, looking at the spectrum in section \ref{secgravity}, and considering the dual forms too, we can write down the spectrum of branes in the theory, as in Table \ref{branes}, together with their proposed YM interpretation, discussed in the following subsections.
\begin{table}[ht]
\caption{Branes in 5d and their proposed YM interpretation.}
\begin{center}
{\begin{tabular}{@{}cccc@{}}\hline
Brane & Field & Sector &$\qquad $ YM interpretation\\
\hline
D(-1) & $\qquad \chi\qquad $ & RR & $\qquad $ instanton\\
\hline
D0 & $\qquad A_1^{R}\qquad $ & RR &$\qquad $ baryon vertex\\
\hline
NS0 & $\qquad A_1^{NS}\qquad $ & NSNS &$\qquad $ monopole vertex\\
\hline
F1 & $\qquad B\qquad $ & NSNS & $\qquad $chromoelectric flux\\
\hline
D1 & $\qquad C_2\qquad $ & RR & $\qquad $chromomagnetic flux\\
\hline
D2 & $\qquad C_3\qquad $ & RR &$\qquad $ domain wall\\
\hline
D3 & $\qquad C_4\qquad $ & RR &$\qquad $ colors\\
\hline
D4 & $\qquad C_5\qquad $ & RR & $\qquad $flavors\\
\hline
\end{tabular}}
\label{branes}
\end{center}
\end{table} 
Actually, the identification of the D3 and D4-branes in the dual filed theory is trivially in terms of color and flavor degrees of freedom support, respectively, as stated in section \ref{secsetup}.
For what concerns the other branes, although the story is similar to the known 10d duals (we will mostly follow the analysis of  \cite{ooguri1,ps}), their identification brings to interesting implications.

\subsubsection{D(-1)-brane: the small instanton}

This is well known from the 10d duals \cite{banks}.
The D(-1)-brane looks like an instanton on the world-volume of the D3-branes.
It couples to the axion $\chi$, whose WZ coupling with the D3-brane is just
\begin{equation}
\int d^4x\  \chi\ F\wedge F\ .
\end{equation}
Thus, the axion provides the YM $\theta_{YM}$ angle, $\chi(r=\infty)=\theta_{YM}$ in the microscopic lagrangian.
We will assume for simplicity that the $\theta_{YM}$ angle is zero in the following.

\subsubsection{D0-brane: the baryon vertex}\label{d0}

The D0-brane is just a particle and couples to the one-form potential $A_1^{R}$.
Fundamental strings (F1) can in principle end on the D0-brane.
Thus, the D0 is a source for the F1 flux, which is naturally associated with the chromoelectric flux.
We propose to interpret the D0-brane as the baryon vertex.\footnote{As pointed out in  \cite{ps}, in the 10d model dual to ${\cal N}=1^*$ theory, the baryon vertex and the domain wall look like D0 and D2-branes in the non-compact five dimensions.}
The open strings coming from the flavor D4-branes would end on it, creating a bound state.

In 10d models the baryon vertex is always a Dp-brane wrapped on an internal compact manifold of dimension p.
Since in the background there is typically a (p-1)-form potential $C_{p-1}\sim N_c$, on the world-volume there is a non-trivial electric field of magnitude $N_c$, which is inconsistent on a compact space unless there are $N_c$ electric sources: the open strings ending on the brane \cite{witten,ps}.

Crucially, there is also a simple picture of this mechanism from the non-compact five dimensional point of view \cite{ooguri1}.
It relies on the observation that the supergravity action contains a CS term that can be reduced to 5d once we integrate on the compact space.
The latter operation gives a factor of $N_c$, so the five dimensional CS coupling contains a factor
\begin{equation}\label{cs5d}
S_{CS} \sim N_c \int d^5x\ \epsilon_{ab}\ B^a \wedge dB^b\ , 
\end{equation}
where $B^a$ is a doublet of (NSNS,RR) two-forms.
But the NSNS two-form $B$ is coupled to the fundamental strings, so a generalization of the standard 3d CS picture implies that one can join $N_c$ fundamental strings together.
In fact, this configuration is gauge equivalent to a zero NS charge configuration, thus overcoming the obstruction to join fundamental strings because of the NS charge conservation.

The same mechanism should hold in the 5d dual to QCD.
The gravity fields comprise both the NSNS and the RR two-forms, and the background includes a non-trivial zero-form field strength $F_0$ dual to $F_5$, so proportional to $N_c$ (this was assumed in the setup in section \ref{secsetup}).
Thus, it is natural to have a CS term for these fields of the form (\ref{cs5d}), where the $N_c$ factor would be given by $F_0$.
If such a CS coupling exists, the same argument as above would imply that one can form a baryon vertex.

Moreover, the D0-branes are point-like instantons on the five dimensional world-volume of the flavor D4-branes.
It has recently been discussed, in the context of the D4/D8 model \cite{sstutti1,sstutti2,sstutti3,sstutti4,sstutti5}, how instantonic configurations in five dimensions reduce to the Skyrme term in four dimensions.\footnote{From  \cite{ckp} we also learn that the WZ term on the flavor brane should couple to $F_0=\ ^{*}F_5\sim N_c$ via $S_{WZ}\sim \int F_0 \cdot \Omega_5$, where $\Omega_5$ gives the five dimensional field theory CS term.}
This strengthens the identification of the D0 as the baryon vertex. 
Unfortunately, our ignorance of the 5d actions prevents a more detailed discussion.  

\subsubsection{NS0-brane: the magnetic monopole vertex}

The one-form potential $A_1^{NS}$ is a magnetic field with respect to the fundamental string.
It couples to a NS0-brane, which is just a particle.
The D1-brane can in principle end on the NS0-brane.
Thus, the NS0 is a source for the D1 flux, which is naturally associated with the chromomagnetic flux.
All these considerations point to the identification of the NS0-brane with a magnetic monopole vertex.

As it is well known, a possible mechanism of confinement in the YM theory is via monopole condensation. 
The latter has the consequence that the monopoles feel no potential: the magnetic charge is screened.
Thus, the 't Hooft loop of the chromomagnetic flux must not give the area law.
This is discuss in section \ref{secD1}.   

The NS0-brane has been dubbed ``magnetic baryon'' in  \cite{gk}, along the lines of  \cite{ps}, as it would be the S-dual of the D0 brane. In fact, since the argument in section \ref{d0} is symmetric in the RR and NSNS 2-forms, it would seem that the D1-branes can be attached to the NS0 only in groups of $N_c$ (this is the reason why we dubbed the NS0 ``vertex''). 
Nevertheless, as discussed in section \ref{secD1}, the monopole screening implies that the D1 are tensionless at the bottom of the space. As such, they are not forced to end on the NS0, which is therefore completely shielded.
It would be interesting to understand better the physics of this object.

\subsubsection{F1: the chromoelectric flux}

This identification is obvious.
The fundamental string is just the YM flux tube.
The fact that it has a finite tension just follows from the requirement that the warp factor in the geometry (\ref{metric}) has a non-vanishing value at the bottom of the space (as we stated in section \ref{secsetup}), since then the Wilson loop gives automatically the area law \cite{wym,kinar}.
In fact, the action for a fundamental string describing a Wilson loop is
\begin{equation}\label{wl}
S \sim \int d^2x\, \sqrt{-G}\ ,
\end{equation}   
where $G$ is the determinant of the induced metric on the world-volume. 
This action is minimized at $r=0$, where $\sqrt{-G} \sim e^{2f(0)}\neq 0$, giving thus a finite tension $\sigma\sim e^{2f(0)}$ and a linear potential in the limit of large separation $L$ of the two test quarks, $V=\sigma L$.
We also recall that the Luscher term comes out trivially for this configuration. 
In fact, four out of the five string world-sheet bosons, representing the fluctuations around the classical configuration, are massless modes to leading order (the boson representing the radius fluctuation has instead a potential proportional to $e^{2f}$ and the world-sheet fermions most probably take mass by the coupling to the RR five-form as it happens in the 10d models). 
Thus, the first correction to the string energy, given by the sum of the zero-point energies of the world-sheet massless modes and of the ghosts, is precisely $-\frac{(n-2)}{24\pi L}=-\frac{1}{12\pi L}$ ($n$ being the number of massless modes).

In $SU(N_c)$ YM the k-strings take values in $Z_{N_c}$ (= center of the group), so one expects $N_c/2$ different strings.
As in  \cite{ps}, the presence of these k-strings is deduced from the presence of the baryon vertex: the k-string is just a bound state of k ordinary strings; if we put $N_c$ of them together, they can pair produce baryon vertexes;
then, the flux tubes can annihilate, showing that a $N_c$-string is equivalent to no string at all, and thus that the strings take charge in $Z_{N_c}$.

Note that the flux tube can break if there are dynamical quarks.
This is trivial in our case: the D4/anti D4-branes are space-time filling, so the F1 live on them and can in principle break with pair production of quarks.
But since there is a chiral condensate and the open string tachyon has a diverging profile in the IR \cite{ckp}, the flavor branes should ``effectively'' end at finite distance from the bottom of the space, so the electric fluxes are metastable: sitting at $r=0$ they can break only if, via world-sheet instanton transitions, they reach the effective bottom of the D4-branes.
Of course, the same is true in the massive quark case, where the radial position of the effective bottom of the branes is a function of the quark mass $m_q$, so the flux tube breaking effect should be exponentially suppressed with $m_q$.

\subsubsection{D1: the chromomagnetic flux}\label{secD1}

This is an obvious identification too.
The important point is that the magnetic flux should not be confined.
This should be reflected in the vanishing of the tension of the D1-brane at the bottom of the space.
Thus, the energy density would be zero: the flux would be dissolved, the D1-brane would ``percolate'' and basically fill up the space.

As it is well known, the tension of the chromomagnetic flux can be derived from the 't Hooft loop.
In 5d, its only difference with respect to the Wilson loop (\ref{wl}) is that the D1 DBI action has a dilaton factor multiplying the area spanned by the string,
\begin{equation}
S \sim \int d^2x\ e^{-\Phi} \sqrt{-G}\ .
\end{equation}
The induced metric on the world-volume is the same as the one for the Wilson loop, which gave 
a finite tension because $\sqrt{-G} \neq 0$.
We conclude that, in order to have magnetic screening, that is a zero tension for the magnetic flux given by the D1-brane, it is necessary and sufficient that the dilaton blows up at the bottom of the space $r=0$, 
\begin{equation}
e^{\Phi} \rightarrow \infty \qquad {\rm for} \qquad r \rightarrow 0\ .
\end{equation}

Now, in this five dimensional D3 model, the YM coupling must be proportional to the dilaton exponential.\footnote{The presence of the closed string tachyon would make no difference in this respect: the same combination of the dilaton with the tachyon would enter both the D1 tension calculation and the YM coupling, so the latter would still blow up.}
Thus, we are led to conclude that the YM coupling must diverge in the far IR. 
This feature is also present in some 10d models (for example in  \cite{wym,min2}).
Thus, it seems that string theory has a tendency of predicting a diverging YM coupling in the IR, rather than a flattening to a constant value.

Let us close this section with a comment on oblique confinement.
At non trivial $\theta_{YM}$, in YM all the charges are confined, apart from the specific cases when $\theta_{YM}=\frac pq$ (with coprime $p,q$), where $(p,q)$-dyons of electric (magnetic) charge $p$ ($q$) condense.
It is well known that in ten dimensional Type IIB theory the tension of a $(p,q)$-string, that is a bound state of $p$ F1s and $q$ D1s, obeys the formula
\begin{equation}\label{oblique}
\sigma_{(p,q)}=\sigma\, e^{-\Phi} \sqrt{e^{2\Phi}(p-\chi q)^2+q^2}\ ,
\end{equation}
where $\sigma$ is the fundamental string tension and $\chi$ is the axion \cite{schwarz2}. 
This formula would give oblique confinement provided the axion is $\chi=p/q=\theta_{YM}$ at $r=0$ (i.e. it does not change along he RG flow, as indicated by the lattice studies in  \cite{Giusti:2001xh,Luscher:2004fu}).
In fact, for generic $\theta_{YM}= \chi$, in the limit $e^{-\Phi}\rightarrow 0$ for $r \rightarrow 0$, formula (\ref{oblique}) gives $\sigma_{(p,q)} = \sigma\, |p-\chi q| \neq 0$, that is, all the charges are confined.
Instead, for $\theta_{YM}= \chi = p/q$, exactly the $(p,q)$-strings have vanishing tension, that is, the $(p,q)$-dyons are screened.
Of course, in five dimensions formula (\ref{oblique}) cannot be the complete story, since the string interactions cannot be discarded.
At best, the formula can approximate the actual behavior for $p,q \ll N_c$.
With a larger amount of strings, the expected behavior should give the ``sine'' or ``Casimir'' scaling of the tension of a $p$-string.
Nevertheless, we expect a similar structure to (\ref{oblique}), with maybe more general functions of $(p-\chi q)$ (for example, a sine), but with the same pattern of dependence on the dilaton, such that the nice behavior described above would not be spoiled.   

\subsubsection{D2: the domain wall} 

The properties of the domain walls, that is that the F1s can end on them and that the baryon vertex can dissolve on their world-volume, are trivially satisfied by the D2-branes (the dissolution of the D0 baryon vertex should be viewed by the WZ coupling $\int A_1^{R}\wedge F_2$, that gives a D0 charge when the gauge field $F_2$ is non-trivial on the world-volume of the D2).
In fact, the D0/D2 system realizes in a natural way the scenario, advocated in  \cite{gs},\footnote{This scenario was ultimately inspired by string theory.} where the domain walls are ``made up'' of soliton-like excitations that resemble baryon vertexes (in the sense that $N_c$ fundamental strings can attach to them and make them escape from the domain wall).\footnote{This was already observed in  \cite{ps}.}

Finally, as for the D1-brane before, a crucial point is the possible vanishing of the D2-brane tension at the bottom of the space, where the branes are dragged by the requirement of minimizing their energy in the background (\ref{metric}). 
Instead, in pure YM at $N_c=\infty$ there are infinitely many stable vacua \cite{Witten:1998uka}.
Even if they are non-degenerate in energy and there exists only one true vacuum (for $\theta_{YM}=0$), all the unstable vacua become stable at infinite $N_c$.
Thus, we expected finite tension domain walls separating these vacua in the model.
It is not clear how this scenario could be realized, but a possibility is that some field is non-trivial on the world-volume of these D-branes, making their tension finite.
For example, in the case the domain walls are really ``made up'' of D0s, a non-trivial $A_1^{R}$ would couple to a magnetic field on the D2, possibly giving it a finite tension, along the lines of  \cite{Schmidhuber:1996fy}.

\section*{Acknowledgments}

We would like to thank A. Andrianov, F. Bigazzi, R. Casero, A. Celi, R. Enparan, U. Gursoy, E. Kiritsis, L. Martucci, C. Nunez, H. Ooguri, A. Paredes and L. Tagliacozzo for useful discussions.
We also thank the referee of IJMPA for pointing out some problems in section \ref{secgravity} in a previous version of these notes.
This work is
partially supported by the European Commission contracts
MRTN-CT-2004-005104, MRTN-CT-2004-503369, CYT-FPA-2004-04582-C02-01,
CIRIT-GC-2001SGR-00065.

\end{document}